# NEED FOR A SOFT DIMENSION

Pradeep Waychal[1] and Luiz Fernando Capretz[2]

[1]College of Engineering, Innovation Center, Pune, India
Pradeep.Waychal@gmail.com
[2]Dept. of Electrical and Computer Engineering, Western University, London, Canada,
lcapretz@uwo.ca

## ABSTRACT

*It is impossible to separate the human factors from software engineering expertise during software development, because software is developed by people and for people. The intangible nature of software has made it a difficult product to successfully create, and an examination of the many reasons for major software system failures show that the reasons for failures eventually come down to human issues. Software developers, immersed as they are in the technological aspect of the product, can quickly learn lessons from technological failures and readily come up with solutions to avoid them in the future, yet they do not learn lessons from human aspects in software engineering. Dealing with human errors is much more difficult for developers and often this aspect is overlooked in the evaluation process as developers move on to issues that they are more comfortable solving. A major reason for this oversight is that software psychology (the softer side) has not developed as extensively*

## 1. INTRODUCTION

The 2004 version of ACM/IEEE Software Engineering Curriculum mentioned document suggests only 5 hours of studies for group dynamics, whereas the 2013 draft recommends 8 hours. This is clearly not enough for a topic of such crucial importance. Besides, there should be more venues to publish papers with results in this field; workshops and conference sessions can help to increase visibility on this area and to discern the connection between human factors (including the individual, social, cultural, and organizational aspects) and the process of software development. But educators and researchers willing to venture into this area should not face an arduous task if they try to convince their colleagues and software engineering "purists" of the importance of the subject. We need to realize that the human element is pivotal to the engineering of software, and it is worth studying and teaching the soft dimension.

Human-Computer Interaction (HCI), a common course within computer science departments, may be the closest subject and great strides have been made by computing professionals in adopting a human viewpoint to improve user interfaces. Although HCI addresses different topics, as it focuses on people interacting with software, HCI designers have been able to add this new dimension to their design philosophy.

Likewise, software engineers could benefit immensely if even the smallest insights of human factors could be incorporated into their way of thinking. My own experiences as a software engineering educator, manager, and practitioner who continually keep human factors in mind. A course on human factor in software engineering should focus on providing a practical overview of the software engineering process from a human factor perspective, an alternative within a panorama of technically saturated curricula.

The 2004 version of ACM/IEEE Software Engineering Curriculum mentioned document suggests only 5 hours of studies for group dynamics, whereas the 2013 draft recommends 8 hours. This is clearly not enough for a topic of such crucial importance. Besides, there should be more venues to publish papers with results in this field; workshops and conference sessions can help to increase visibility on this area and to discern the connection between human factors (including the individual, social, cultural, and organizational aspects) and the process of software development.

However, educators and researchers willing to venture into this area should not face an arduous task if they try to convince their colleagues and software engineering "purists" of the importance of the subject. We need to realize that the human element is pivotal to the engineering of software, and it is worth studying and teaching the soft dimension.

## 2. ACCREDITATION ATTRIBUTES

Ample work has been reported on the required attributes of software engineering graduates and serious gaps between them and available ones. All the accreditation bodies have cognized that and included a set of attributes that the graduating engineers must have. The specific attributes that are in the limelight are teamwork, critical and creative thinking, ethics, lifelong learning, and communication.

While professional organizations have some qualitative means to evaluate these skills and requisite development programs, the academic environment neither measures nor develops them. This needs to change. We certainly need a systemic approach in both industry and academic environments. That entails a regular cycle of measurement, conceptualization, and execution of development programs. Given the academic load on the students, it is imperative that the development of the attributes is integrated with the core engineering curriculum as recommended by Honor [1]. We will describe our experience in the next few sections.

### 2.1. Teamwork

Teamwork is involved in virtually every professional activity and therefore should be embedded in every possible academic activity. Colleges and Universities should attempt to bring in as much diversity – in terms of academic performance, social and linguistic background, discipline, gender and culture – as possible in their student teams. They should endeavour to assign different roles to students in different teams so that they can develop varied skills. They should also provide the experience of having interactions across cultures and across disciplines both in physical and virtual modes.

The teamwork skill can be measured using instrument developed by Ohland *et al*. [2] and the one based on "Ten Commandments of Egoless Programming" proposed by Adams [3]. We have been using them and appropriate instructional strategies resulting in students getting immense benefits.

### 2.2. Creative and Critical Thinking

The engineering field is becoming increasingly complex across all its branches - from traditional civil engineering to modern computer, software, space, mechatronics and genetic engineering. The complexity has increased even more due to a growing interdependence among disciplines and the emergence of a wide range of new technologies. To manage this situation, engineers who are creative and capable of abstract thinking, engineers who can keep pace with new technologies and think laterally when developing new applications are needed. It has been observed that the recent engineering graduates are lacking in these competencies [4]; and the traditional and still dominant engineering curriculum at most universities, especially in developing countries, makes little provision for developing these competencies [5].

We have been experimenting with Index of Learning Style (ILS)'s [6] sensing intuition preferences to measure critical thinking. We have chosen this instrument over other options like MBTI and TTCT since the latter are costly and have elicited diverse opinions on their suitability. We have also designed a course and found statistically significant changes in critical thinking based on the ILS measure.

## 2.3. Ethics

This is a complex skill to measure and develop. Its development requires involvement of other stakeholders like parents, society, and industry. Right now, academicians are handling it by introducing a separate traditional or scenario based course. That is not proving to be sufficient as it lacks real life situations with real software engineering stake holders [7]. We have introduced peer evaluation using constant sum scale for many courses and believe correlations in the self and peer evaluation may provide some idea about ethics of students.

## 2.4. Life Long Learning

In today's knowledge economy continuous/lifelong learning is assuming greater significance. Every educational institute needs to appreciate that and make provision for developing those skills, for instance to nurture the ability to understand available code written in modern languages in open source software libraries [8].

Colleges require introducing courses or at least some topics in the courses where students have to learn on their own. Performance in them may indicate their skill in the attribute. We have introduced courses on Liberal Learning for all students at the College of Engineering Pune, India. They study various non engineering topics on their own. That provides opportunities to develop lifelong learning skills and a method to evaluate their performance in that important dimension [9].

## 2.5. Communication

This is critical not only for the engineering career but for all careers. It does not just mean knowledge of English and basic verbal skills but also includes the ability to write emails and proposals, articulate and communicate ideas, create and deliver public presentations, interact across cultures, and listen and analyze talks. This is the easiest of the skills and needs to be developed and measured in every academic course.

For example, engineering software design involves performing tasks in distinct areas, such as system analysis, software design, programming, software testing, and software evolution/maintenance [10]; other software occupations on a design team include the project manager, troubleshooter, helpdesk personnel, database administrator, and so forth. Thus today, specialties within software engineering are as diverse as in any other profession [11]. Therefore, software engineers need to communicate very effectively with users and team members, consequently the people dimension of software engineering is as important as the technical expertise [12].

## 3. CONCLUSIONS

Software engineering has been doing a marvellous job of helping enterprises of all types and to continue doing so it requires focusing on the soft dimension – people dimension. Its development must start in the undergraduate and graduate courses. The primary areas to work on are: teamwork, creative and critical thinking, ethics, lifelong learning and communication. They still need to be both developed and measured.

**AUTHORS**

**Pradeep Waychal** chairs an NGO that works at the intersection of human sciences and software engineering, engineering education and innovation management. He has done Ph.D. in developinginnovation competencies for Information System Businesses from IIT Bombay, India. He is a senior member of IEEE, a member of ASEE and a life member of CSI – India.

**Luiz Fernando Capretz** is a professor of software engineering and assistant dean (IT & e-Learning) at Western University in Canada, where he also directed a fully accredited software engineering program. He has vast experience in the engineering of software and is a licensed professional engineer in Ontario. Contact him at lcapretz@uwo.ca or via http://eng.uwo.ca/electrical/faculty/capretz_l/